# Social Impact of MOOC's in Oman Higher Education

Dr. RD.Balaji,
College of Applied Sciences - Salalah,
Sultanate of Oman,
balajicas@yahoo.com,

Mrs. Fatma Al-Mahri,
College of Applied Sciences - Salalah,
Sultanate of Oman,
fatmam.sal@cas.edu.om

Mr. Tarek Al-Fatnassi
College of Applied Sciences - Salalah,
Sultanate of Oman,
tarek.sal@cas.edu.om

**ABSTRACT**

The word "E" transformed everything is this world, as well as the whole globe itself. To a great extend this helps for eco friendly green world. In educational field, electronic medium has played a major role. It influenced and changed almost every component of it to electronic medium like e-book, online courses, etc. Throughout the world, leading universities are offering online courses voluntarily. Generally we refer to it as Massive Online Open Courses (MOOCs). There are many debates going on related to success and consequences of MOOCs. Many are highlighting that these courses are self-paced, economical, and provide quality training to all irrespective of geographical constraints. But many other academic people go against these points and keep listing many other disadvantages of MOOCs. This paper explores the basics of MOOCs at the initial section. Following section will deal with advantages and disadvantages of MOOCs in general. We the researchers collected the details about the awareness of MOOCs among teachers and students in a higher education institution in Oman. We have also collected the details about MOOCs implementation and usage within Oman educational society. Based on the collected information, we have evaluated and presented the findings about MOOCs impact in Oman higher education. We have felt that doing appropriate improvements in MOOCs may become an imperative medium in Oman educational institutions. The suggestions are listed in the discussion and recommendation section.

**KEYWORDS**

MOOC, ICT, Higher Education, Flipped Classroom, M-Learning, E-Learning

**INTRODUCTION**

All the countries in the world understood that the education is the only way to take the society to the next level. But in recent days cost of educational service increased in many folds. More than 80 percent of the world population is refused to access the quality education. The top most universities in the world have taken this as a challenge and wish for providing quality education to large group of students with no limits of access, attendance and credits offered for the courses. This concept is globally accepted as MOOC. This acronym was first used during December 2011 in reference to an artificial intelligence course offered by a Stanford University professor [1]. Even though this concept is introduced few years before, it is well received by the society throughout the world including students from developing and under developed countries [2]. MOOC takes the advantages of e-learning, m-learning, flipped classroom and multimedia technology, provides quality education without much hazel to huge number of self-interested students and does not hold any geographical constraints. Even in the short span of time from the evolution of this concept, creation and dissemination of high-quality digital lecture materials benefit both online and on-campus students [3]. Even though MOOCs have gained lot of popularity in this short period of time, there are strong debates going on about its limitations. The main advantages and disadvantages will be discussed in the next section. We will also discuss about the characteristics of MOOCs in the following section.

**MOOC**

MOOC is the acronym of Massive Open Online Course. There are many companies and universities, which are offering MOOC courses. Coursera, The Open University, Iversity, Edx, Edukart are the few popular MOOC curators. Most of the companies are started by the educationalist and also with non-profitable motive. The name itself directly reflects the characteristics of MOOC. The courses are offered for Massive crowd, openly and following Collaborative philosophy. Thrun and Norvig's Artificial Intelligence course (CS221) was registered by 160,000 students from 190 countries. Later the same professor launched Udacity in 2012 [4]. "*From a pragmatic perspective, MOOCs provide access to large numbers of people who might otherwise be excluded for reasons ranging from time, to geographic location, to formal prerequisites, to financial hardship*" [5]. In most of the courses in MOOC, the software, curriculum, assessments, discussions, and registration are open to everyone at free of cost [6]. In MOOCs, only the self-interested, self-motivated students can be successful in the completion of courses. Hence the students who enroll course in MOOC must self-organize themselves and participate in a course where it matches with their prior knowledge, skills and learning goals. Due to the massive nature of MOOC it is not possible for the lecturer to interact with all the students. Hence it is student's responsibility to engage themselves with the course activities and interaction with their peers.

The limitations of MOOCs are mainly pointed out by higher education institute faculty, administrators and trustees. The main limitation pointed out by all these people is the high dropout rates in MOOCs courses [7]. Based on this point, faculty of higher education argues that MOOCs will de-skill and disinter mediate faculty, thereby compromising the quality of education and students experience [8]. Even students are not having full confidence about the output of MOOC courses since it is not recognized by many universities and the companies. But if MOOC becomes successful, then the economy of a country may get affected, since it may lead to massive withdrawal of public funds for higher education [9]. Irrespective of these disadvantages, MOOC is growing exponentially, because it seems to be a "silver bullet" for financial pressures.

MOOC may replace all the current form of traditional teaching and learning, if it is supported by the Governments, universities and the corporate people. The policies need to be formulated where there is an ambiguity in the implementation and acceptance of MOOC.

**RESEARCH METHODOLOGY**

This research has been conducted as a case study and it has utilized few methods for collecting and analyzing the collected data [10]. Since this research is conducted within a college and with all teaching staff members, we have adopted mixed-method approach that combined quantitative and qualitative data collection [11]. The main quantitative collection has been done by the questionnaire filled in by all the academic staff members who are working in the College of Applied Sciences, Salalah, the Sultanate of Oman. A total number of 78 staff members filled the questionnaire. This provided valuable information on staff profile, their proficiency about MOOC and details about their experience in MOOC courses. We also have tried to implement the same method for collecting the data from the students. Before circulating the questionnaire, through grapevine communication method we have found that almost all the students are not aware of MOOC either or they haven't studied any course in MOOC. Hence we have decided not to extend this research to students.

**CAS STAFF PROFILE**

We, the research team wanted to conduct this research at the intra college level first before extending this to other colleges. The sample we have taken is all the teaching staff members of the College of Applied Sciences. All the Staff members are recruited through Ministry of Higher Education. Around 30 percent of the staff members are Omani's and others are expatriates. The minimum qualification is master degree. The minimum experience required to work in the ministry is five years for expatriates. Omani's can join without five years experience as assistant lecturers. We have included all the teaching members in this research. The department wise numbers of staff details are given below in table 1.

| Sl.No | Department Name | No. of Staff |
|---|---|---|
| 1. | Information Technology | 15 |
| 2. | Communication | 9 |
| 3. | Business | 15 |
| 4. | English | 35 |
| 5. | General Requirement | 4 |
|  | Total | 78 |

Table 1: Department wise Staff details

There are 19 female staff and 59 male staff members in the Salalah College of Applied Sciences. Since we are concentrating on general awareness of MOOC among all the academic staff, we have ignored the age and years of experience of each faculty members. But we decided to take the account of positions held by the teaching staff. We have grouped the lecturers and assistant lectures together. There are 20 assistant professors and 58 lecturers working in the college. Since there are very few students having awareness of MOOC, we did not provide the students profile in this section. In the department of Information Technology and Business, there are considerable amount of staff members having awareness of MOOC. The percentage of faculty members aware of MOOC in these two departments is 86.66 percent (26 out of 30). Apart from these two departments, when we calculate the awareness of MOOC in all other departments, it comes around only 18.75 percent (9 out of 48) only. The further details are discussed and the recommendations based on these details are explained in the next section.

**DATA ANALYSIS AND RECOMMENDATIONS**

In the previous section we have discussed the profile of the teaching staff in Salalah College of Applied Sciences. The profile of the staff members shows that it is completely heterogeneous [12]. Since MOOCs are largely aimed at the population outside the formal education, our sample perfectly fits with it. All the faculty members are between the age group of 25 and 55 years which is not coinciding with the digital native age group, which makes people feel difficult to work with MOOC due to technical difficulties. Hence the derived results from our data collection are little deviated with the international scenario. The table 2 shows the details about the number of staff members who are all having awareness of MOOC, number of staff members taking MOOC courses and number of people not aware of MOOC at all.

|  | Number of Staff Members |
|---|---|
| Only aware of MOOC | 17 |
| Taking course with MOOC | 18 |
| Not aware of MOOC | 43 |

Table 2: Awareness of MOOC among staff members

The staff members who are aware of MOOC and not taking any courses have listed the reason for not taking the courses in the questionnaire. The top 5 reasons are listed below after summarizing the collected data. Similarly there are some staff withdraw the registered course in the middle due to various reasons. There are totally 62 courses registered by the teaching staff of Salalah College of Applied Sciences (hereafter Salalah CAS), but they could complete only 37 courses. All other courses are withdrawn by the staff members in the middle due to various reasons. These reasons are also combined with the reasons for not registering courses in MOOC and listed below. The failure rate of MOOC courses are around 40.323 percent.

1. **Free of Cost**: The main objective of MOOC is to provide quality education at free of cost. But when a student register at free of cost, they don't have commitment towards it. When they encounter some other urgent work, they leave this course in the middle. This is the top most reason for a student to dropout in the middle of the course.

**2. Irrelevant**: Majority of the time the course title is misunderstood by the student. Many people felt that the course is not what they thought it was, after attending the first week class or few week classes of the course and discontinue in the middle.

3. **Technical Difficulties**: MOOC courses are completely offered and evaluated online, hence the assessment form is not like the traditional one. One who is well versed in many components of the internet like using YouTube, Facebook, recording videos, using webcam etc., only can perform well in the MOOC course. The students, who may be strong in their subject but not expert in web tools, are also forced to withdraw from the course.

4. **Language Barrier**: For the people, those who want the courses in their Mother tongue like Arabic, very few (sometimes null) courses are offered in MOOC. Hence many are discouraged to register course through MOOC.

5. **No Recognition**: Students can register course only for their knowledge. MOOC course certificates are not recognized in the institutions, as well as companies in Oman. Hence students are not interested in doing course for their carrier advancement or new job opportunities.

Apart from the above mentioned reasons many are feeling that the work load in college is not encouraging them to go for any additional course. People are finding it difficult to find a very suitable course which may help them in handling courses in Oman. Another major reason is general technical difficulties in taking MOOC courses and language issues as discussed above.

Interesting observation made by the research team is derived from the table 3.

| Sl. No | Course Numbers | Number of Staff |
|---|---|---|
| 1. | < 2 | 11 |
| 2. | 3 to 5 | 5 |
| 3. | Above 5 | 2 |

Table 3: Number of courses registered by staff members

The above table shows that very few are doing more courses with MOOC. This is directly related with the initial success of completing the courses. If someone is able to complete the course, then they are interested in continuously registering new courses with MOOC. If not, they have a lot of hesitation in registering any new courses. This is again proves that technical compatibility is very much required to complete their course in any domain. They need time to do the course during the semester; if they are not able to manage their time then the willingness rating to register new courses with MOOC goes down.

| Sl. No | Course Rating | Number of Staff |
|---|---|---|
| 1. | Excellent | 7 |
| 2. | Good | 12 |
| 3. | Average | 20 |
| 4. | Poor | 15 |

Table 4: Rating the MOOC courses by Number of Staff

When we want to know the staff rating of MOOC courses, many have rated MOOC even though they have not taken any courses with MOOC. Hence the reliability of this data is uncertain. But by personal interview the research team found that they have viewed few courses in MOOC and rated the course without taking any course.

The general opinion by all the staff members is that all are willing to take the courses with MOOC if it is encouraged by the college and recognized by the ministry, provided they also need suitable timings and workload so that they can take MOOC courses. More than 90 percent of the staff members are interested in doing Course with MOOC due to the quality of the courses.

These are the observations made by the research team and we are seeing promising future for MOOC at the teacher's level soon. It may take lot of time to have impact with students unless otherwise the ministry takes policy decisions in encouraging MOOC.

**CONCLUSION**

This paper has focused on the impact of MOOCs in the higher educational institutions in the Sultanate of Oman. The students and the teaching fraternity have been involved in this research. Unfortunately, in the beginning of the research, authors found that the awareness of MOOCs among post diploma students is even less than 5 percentages. Hence the complete research was conducted among the teaching staff members to find out how much MOOC influenced them and what will be the future of MOOCs in the higher education institutions. Since the advertisements of the MOOCs are not focusing on the right group of people, many potential related group are not aware of MOOCs. In this paper we are not discussing about the international impact of MOOC and other attributes of MOOC. Our main objective is to find out in Oman higher education institute how far MOOC is influencing the stakeholders. The educational ministry and technological associations in Oman should take more initiatives to popularize MOOCs among the teaching community. There is no policy decision taken by the Government about MOOC as part of the formal education structure of Oman's higher education. Generally, the pressure in the formal education is not encouraging the students to take value added courses through MOOCs. Awareness on MOOC is very poor among all other teaching staffs except business and IT departments. Even with IT and business staff, only 50 percent of them are doing or did courses with MOOC due to more work load in the college. The research has explored that the infrastructure of the college is excellent for taking courses with MOOC. The staff members who are familiar with MOOC accept that it provides good platform to upgrade their knowledge and stay in touch with the recent development in their domain with nominal cost or for free. The outcomes of this research show promising effects among the higher education institutions even without much effort of the stakeholders. MOOCs will become more powerful when the Government, educational ministries and higher educational institute managements take collaborative initiative to promote MOOC among students and staff members for their benefits with less financial implications.

**REFERENCES**

[1] J, Kim. (2011). Grad Students and Digital Education, Retrieved on December 10, 2014, from http://www.insidehighered.com/blogs/technology-and-learning/grad-students-and-digital-education

[2] Martin-Monje, E., E.Barcena & T.Read (2013) Exploring the affordances of Massive Open Online Courses on second languages. Proceedings of UNED-ICDE (International Council for Open and Distance Education), Madrid:UNED

[3] D.C.Schmidt and Z.McCormick, Creating and teaching a MOOC on Pattern-Oriented Software Architecture for Concurrent and Networked Software, Proceedings of the WaveFront Forum at the SPLASH 2013 conference, October 2013, Indianapolis, IN

[4] Carr, N. (2012, September 27). The crisis in higher education. MIT Technology Review. Retrieved from http://www.technologyreview.com/featuredstory/429376/the-crisis-inhigher-education/

[5] McAuley, A., Stewart, B., Siemens, G., & Cormier, D. (2010). The MOOC model for digital practice, 1-63. Retrieved from http://www.elearnspace.org/Articles/MOOC_Final.pdf

[6] Rodriguez, C. O. (2012). MOOCs and the AI-Stanford like courses: Two successful and distinct course formats for massive open online courses. European Journal of Open, Distance and E-Learning. Retrieved from http://www.eric.ed.gov/PDFS/EJ982976.pdf

[7] Tamar Lewin, "After Setbacks, Online Courses Are Rethought," New York Times, December 10th, 2013.

[8] Rees, "The MOOC Racket," Slate, July 25th, 2013.

[9] Rosenberger, "John L. Hennessy on 'The Coming Tsunami in Educational Technology'," Communications of the ACM Blog, July 23, 2012.

[10] Cohen, L., L.Manion, L & K. Morrison (2007) *Research Methods in Education.* London: Routledge.

[11] Robson, C. (2002) Real World Research: A Resource for Social Scientists and Practitioner-Researchers. Blackwell Publishing, Oxford.

[12] Worlock, K. & L. Ricci (2013) MOOCs: Cutting Through the Hype. http://www.smarthighered.com/wp-content/uploads/2013/04/MOOCs-CUTTING-THOUGH-THE-HYPE.pdf (last accessed: 1/10/2014).